\documentclass[10pt,aps,prl,showpacs,superscriptaddress,nofootinbib,twocolumn,preprintnumbers,floatfix,showkeys]{revtex4-1}
\usepackage{graphicx,amsmath,amssymb,soul,enumitem,bbm,dcolumn}
\usepackage{booktabs,multirow,array,slashed,dsfont,kantlipsum}
\usepackage[dvipsnames]{xcolor}
\usepackage{nicefrac}
\usepackage[utf8]{inputenc}
\allowdisplaybreaks
\usepackage{hyperref}
\hypersetup{
colorlinks=true,
urlcolor=PineGreen,
linkcolor=PineGreen,
citecolor=PineGreen}
\usepackage{enumitem}
\setlist{topsep=1pt}
\usepackage{booktabs}

\newcommand{\aw}{\mathbf{\alpha_-}}
\newcommand{\obs}[1]{{{#1}}}

\begin{document}

\title{Kaon Photoproduction and the ${\boldmath\Lambda}$ Decay Parameter
$\boldmath{\aw}$}

\author{D.\,G.\ Ireland}
\email[Corresponding author: ]{David.Ireland@glasgow.ac.uk}
\affiliation{%
SUPA, School of Physics and Astronomy, 
University of Glasgow G12 8QQ, 
United Kingdom
}%
\author{M.\ D\"oring}
\email{doring@gwu.edu}
\affiliation{Institute for Nuclear Studies and Department of Physics, The George Washington University, Washington, DC 20052, USA}
\affiliation{Thomas Jefferson National Accelerator Facility, Newport News, VA 23606, USA}
\author{D.\,I.\ Glazier}
\email{Derek.Glazier@glasgow.ac.uk}
\affiliation{%
SUPA, School of Physics and Astronomy, 
University of Glasgow G12 8QQ, 
United Kingdom
}%
\author{J.\ Haidenbauer}
\email{j.haidenbauer@fz-juelich.de}
\affiliation{Institute for Advanced Simulation, Institut f\"ur Kernphysik (Theorie) and J\"ulich Center for Hadron Physics, Forschungszentrum
J\"ulich, D-52425 J\"ulich, Germany}
\author{M.\ Mai}
\email{maximmai@gwu.edu}
\affiliation{Institute for Nuclear Studies and Department of Physics, The George Washington University, Washington, DC 20052, USA}
\author{R.\ Murray-Smith}
\email{Roderick.Murray-Smith@glasgow.ac.uk}
\affiliation{%
School of Computing Science, 
University of Glasgow G12 8RZ, 
United Kingdom
}%
\author{D.\ R\"onchen}
\email{roenchen@hiskp.uni-bonn.de}
\affiliation{Helmholtz-Institut  f\"ur  Strahlen-  und  Kernphysik  (Theorie)  and  Bethe
Center  for  Theoretical  Physics,  Universit\"at  Bonn,  53115  Bonn,  Germany}


\begin{abstract}

The weak decay parameter $\alpha_-$ of the $\Lambda$ is an important quantity for  the extraction of polarization observables in various experiments. Moreover, in combination with $\alpha_+$ from $\bar\Lambda$ decay
it provides a measure for matter-antimatter asymmetry. The weak decay parameter also affects the decay parameters of the $\Xi$ and $\Omega$ baryons and, in general, any quantity in which the polarization of the $\Lambda$ is relevant. The recently reported value by the BESIII collaboration of $0.750(9)(4)$ is significantly larger than the previous PDG value of $0.642(13)$ that had been accepted and used for over 40 years.
In this work we make an independent estimate of $\alpha_-$, using an extensive set of polarization data measured in kaon photoproduction in the baryon resonance region and constraints set by spin algebra. The obtained value is 0.721(6)(5). The result is corroborated by multiple statistical tests as well as a modern phenomenological model, showing that our new value yields the best description of the data in question. Our analysis supports the new BESIII finding that $\aw$ is significantly larger than the previous PDG value.  Any experimental quantity relying on the value of $\alpha_-$ should therefore be re-considered.
%

\end{abstract}



\maketitle


\section{Introduction}

The decay parameter $\aw$ of the parity-violating weak decay $\Lambda \rightarrow p \pi^-$ describes the interference between parity-violating $s$- and parity conserving $p$-waves.
A recent study by the BESIII collaboration~\cite{Ablikim:2018zay} reported a value of $\aw$ as $0.750 \pm 0.009 \pm 0.004$ for this quantity, 
which is significantly different compared to the older value of
$0.642 \pm 0.013$ quoted in the reviews of the Particle 
Data Group (PDG) until 2018 \cite{Tanabashi:2018oca}.

This newly published value of $\aw$~\cite{Ablikim:2018zay} is some 17\,\% higher than the older average PDG value, which had been derived from results in  Refs.~\cite{Astbury:1975hn, Cleland:1972fa} and others, that were not compatible among themselves. Since the BESIII and older average PDG values have uncertainties at the percent level, there is a discrepancy of about five standard deviations, and the two results are therefore incompatible. The discrepancy might be due for instance to underestimated systematic effects in the calculation of correction factors in Ref.~\cite{Astbury:1975hn}. In the case of Ref.~\cite{Cleland:1972fa} photographs of carbon-plate spark chambers were used, and a ten-parameter kinematic fit applied to each event; several sources of uncertainty were highlighted and together with the approximate fitting method, there was ample scope for systematic error. Whilst the previous measurements were all state-of-the-art when carried out, the 2019 PDG online update lists only the new BESIII value ``above the line''.

An independent
estimate of this quantity is highly desirable given that $\aw$ plays
an important role in various fields of physics. 
For instance, comparing $\aw$ with the parameter 
$\alpha_+$ of the decay $\Bar{\Lambda} \rightarrow \Bar{p} \pi^+$
provides a test of $CP$ symmetry for strange baryons and, thus can potentially shed light on the matter-antimatter asymmetry in the
universe 
\cite{Sakharov:1967dj}. 
In this respect, a CP violation at the 3.3 $ \sigma$ level has been found by the LHCb collaboration in four-body decays of $\Lambda_b^0$ and $\bar\Lambda_b^0$ baryons~\cite{Aaij:2016cla}. In the BESIII simultaneous measurement of $\alpha_-$ and $\alpha_+$ of the $\Lambda$, no sign of CP violation was found~\cite{Ablikim:2018zay}, thereby resolving tensions between older PDG values for them. The parameter
$\aw$ has also an impact on several theoretical studies
where its actual value enters directly. In particular, it would
affect calculations of the weak nonleptonic hyperon decays within SU(3)
chiral perturbation theory \cite{Bijnens:1985kj,Holstein:2000yf,Borasoy:2003rc}.

Over the last 40 years there have been various experiments
whose results rely on the value of $\aw$. Examples of this are the
extensive studies of the reactions $\bar pp \to \bar \Lambda \Lambda$
and $\bar pp \to \bar \Lambda \Sigma^0 + {\rm c.c.}$ by the
PS185 Collaboration at the LEAR facility at CERN \cite{Klempt:2002ap} that measured analyzing powers, spin-correlation parameters and spin-transfer coefficients. Recent results, such as the STAR measurement of heavy ion collisions to study the vortical structure of a nearly ideal liquid \cite{the_star_collaboration_global_2017}, and the ATLAS  measurement of $\Lambda$ and $\Bar{\Lambda}$ transverse polarization \cite{atlas_collaboration_measurement_2015}  also depend on the value of $\aw$.

Information about other strange baryons 
depends on $\aw$ through chains of successive decays. For example, the decay parameter for $\Xi$ is determined from
the decays $\Xi \to \Lambda \pi \to N\pi\pi$ and deduced
from the product $\alpha_\Xi \aw$, which in turn affects the measured polarization data for the reactions
$K^-p \to K^+ \Xi^-, K^0 \Xi^0$ \cite{Trippe:1967wat,Dauber:1969hg}
and $\gamma p \to K^+K^+ \Xi^-$ \cite{Bono:2018ike}. The decay parameter for $\Omega^-$ depends likewise on the values
of $\alpha_\Xi$, and therefore $\aw$ \cite{Tanabashi:2018oca}.

Another class of experiments that depends on $\aw$ is the series of measurements of recoil polarization observables for kaon photo- and electro-production in the baryon resonance region \cite{carman_first_2003,carman_beam-recoil_2009,mccracken_differential_2010,bradford_first_2007,paterson_photoproduction_2016}. Up to now, all recoil polarization observables relying on the weak decay of the $\Lambda$ have been evaluated using the pre-2019 PDG value of $\alpha_-$ (henceforth denoted $\alpha_-^{\rm old}$). Fits to such observables by theoretical models are a crucial element in determining the light baryon resonance spectrum~\cite{Kamano:2016bgm,Anisovich:2017bsk,Ronchen:2018ury,Hunt:2018mrt}, which provides a point of comparison for theoretical approaches such as quark models, Dyson-Schwinger or Lattice QCD calculations. 

Kaon photoproduction data can be also utilized to provide a new and independent
estimate for $\aw$, as will be demonstrated in the present work.  
The photoproduction data set contained in the combination of publications \cite{bradford_first_2007,mccracken_differential_2010,paterson_photoproduction_2016} by the CLAS collaboration, is subject to strict constraints from spin algebra (so-called Fierz identities), which can be exploited to derive estimators for $\aw$ itself.
We note that a similar strategy has been followed once before, based
on data for the reaction $\pi^-p \to K^0\Lambda$ \cite{Astbury:1975hn}. 
Anticipating our result, the value for $\aw$ found in our analysis is 
$0.721\pm 0.006$, i.e. close to but noticeably smaller than the 
number given by the BESIII collaboration~\cite{Ablikim:2018zay}. 


\section{Determination of $\alpha_-$ from Kaon Photoproduction data}
\label{sec:formalism}

Photoproduction experiments measure events in bins of hadronic mass $W$, or equivalently Mandelstam $\sqrt{s}$, and center of mass meson scattering angle $\cos\theta.$
Following Ref.~\cite{Sandorfi:2010uv}, the relative intensity distributions of events in each $\{W,\cos\theta\}$ bin for $\gamma + p \rightarrow K + \Lambda$ reactions in which there is no polarization of the beam or target, but where the decay products of the $\Lambda$ are measured, is 
\begin{equation}
    1 + \aw \cos\theta_y \obs{P}.
\label{eq:UUY}
\end{equation}
If the photon beam is circularly polarized we have
\begin{equation}
    1 + \aw \cos\theta_y \obs{P} +
\left(
\aw \cos\theta_x \obs{C_x} + \aw \cos \theta_z \obs{C_z} 
\right) P_C^{\gamma},
\label{eq:CUY}
\end{equation}
and if the photon beam is linearly polarized the distribution is
\begin{equation}
    \begin{aligned}
& 1 + \aw \cos\theta_y \obs{P} -
                        \left\{
                        \obs{\Sigma} + \aw \cos\theta_y \obs{T} 
                        \right\} P_L^{\gamma} \cos 2\phi \\
& \; -
                        \left\{
                        \aw \cos\theta_x \obs{O_x}  + \aw \cos\theta_z \obs{O_z} 
                        \right\} P_L^{\gamma} \sin 2\phi.
\end{aligned}
\label{eq:LUY}
\end{equation}
The $O_j\in\{O_x, \,O_z,\,T,\,C_x,\,C_z,\,\Sigma,\,P\}$ represent the polarization observables and $\phi$ is the angle between the reaction plane and the photon polarization axis. The coordinate system employed in this analysis is the so-called ``unprimed'' frame where, for a photon momentum $\vec{k}$ and a kaon momentum $\vec{q}$, axes are defined such that
\[
\hat{z}=\frac{\vec{k}}{|\vec{k}|};\quad\hat{y}=\frac{\vec{k}\times\vec{q}}{|\vec{k}\times\vec{q}|};\quad\hat{x}=\hat{y}\times\hat{z}.
\] 
The reaction plane is thus defined by the vector $\vec{k}\times\vec{q}$, and the coordinate system attached to the $\Lambda$ at rest uses the same orientation for determining direction cosines of the decay proton $\cos\theta_{x,y,z}$.
Together with $\aw$, the degrees of circular and linear polarizations, $P_L^{\gamma}$ and $P_C^{\gamma}$, enter as ``calibration'' parameters. The three expressions (\ref{eq:UUY}), (\ref{eq:CUY}) and (\ref{eq:LUY}) represent the measurements \cite{mccracken_differential_2010}, \cite{bradford_first_2007} and \cite{paterson_photoproduction_2016} respectively.



Assuming that the angles $\theta_{x,y,z},\phi$ are measured accurately, the extraction of the polarization observables $O_j$ is possible only if the calibration parameters $\{\alpha_-,\,P_C^{\gamma},\,P_L^{\gamma}\}$ are known. The equations (\ref{eq:UUY}), (\ref{eq:CUY}) and (\ref{eq:LUY}), show that the extraction of $O_x$, $O_z$ and $T$ requires the product $\alpha_- P_L^{\gamma}$, $C_x$ and $C_z$ require $\alpha_- P_C^{\gamma}$, whilst $\Sigma$ and $P$ require $P_L^{\gamma}$ and $\alpha_-$, respectively.

The spin algebra of pseudoscalar meson photoproduction results in several constraints among all 15 polarization observables, known as Fierz identities after the method used in \cite{chiang_completeness_1997} to derive them. Two of these connect the observables measured by the CLAS collaboration:
\begin{align}
O^2_x + O^2_z + C^2_x + C^2_z + \Sigma^2 - T^2 + P^2 &= 1 
\label{eq:Fierz1}
\\
\Sigma P - C_x O_z + C_z O_x - T &= 0 \ .
\label{eq:Fierz2}
\end{align}
If all observables in equations (\ref{eq:Fierz1}) and (\ref{eq:Fierz2}) are measured then these Fierz identities can be used to estimate the calibration parameters. The published experiments estimate the uncertainties in $P_C^{\gamma}$ and $P_L^{\gamma}$ as systematic uncertainties, so we have some prior knowledge of their values, giving the opportunity to estimate $\alpha_-$.

The CLAS data span a range of energies $W$ and scattering angles $\theta$. Distributions of observables in $\{W,\cos\theta\}$ are then used to study light baryon resonances. In the present work, we can simply treat the measured data as an ensemble of observations, each of which are related to $\alpha_-$. 

There is a common region in $\{W,\cos\theta\}$ space among the three measurements \cite{mccracken_differential_2010}, \cite{bradford_first_2007} and \cite{paterson_photoproduction_2016}, which is spanned by the 314 points reported in \cite{paterson_photoproduction_2016}. Denoting by 
$O_{j,i} \equiv O_j(W_i,\cos\theta_i)$ the seven observables $j=1,\dots,7$ at kinematic points $i \equiv \{W_i,\cos\theta_i\}$, we have five of these observables, $\{O_x, \,O_z,\,T,\,\Sigma,\,P\}_i;\,i=1,...,314$, from Ref.~\cite{paterson_photoproduction_2016}. 
To obtain the values of $C_x$ and $C_z$ (and their variances) at the points $\{W_i,\cos\theta_i\}$ we proceed as
follows:
We use Gaussian process prior (GP) inference \cite{RasWil05} with maximum a posteriori (MAP) optimisation of covariance function hyperparameters to model the $C_x, C_z$ observation uncertainties. A second heteroscedastic GP is used, incorporating the mean of the GP uncertainty model as observation variance, to interpolate the data reported in \cite{bradford_first_2007}, using the GPML package \cite{RasNic10}.
Illustration and cross-checks of the method are provided in the Supplemental Material~\cite{supplemental}.

\subsection{Statistical Analysis \label{sec:stat}}
With these data, the following Fierz values can be defined:
\begin{align}
{\cal F}^{(1)}_i
=
&a^2 l^2\left({\cal O}_{x,i}^2+{\cal O}_{z,i}^2-{\cal T}_i^2\right) \nonumber \\
&+a^2c^2\left({\cal C}_{x,i}^2+{\cal C}_{z,i}^2\right)
+l^2\mathit{\Sigma}_i^2+a^2{\cal P}_i^2 \ , 
\label{eq:fierzval}
\\
{\cal F}^{(2)}_i
=&al\left[
\mathit{\Sigma}_i {\cal P}_i - {\cal T}_i - a c ({\cal C}_{x,i} {\cal O}_{z,i} - {\cal C}_{z,i} {\cal O}_{x,i})\right] \ ,
\label{eq:fierzval2}
\end{align}
where $c\,(=P_C^{\gamma{\rm old}}/P_C^{\gamma})$ and $l\,(=P_L^{\gamma{\rm old}}/P_L^{\gamma})$ represent relative systematic correction factors in the calibration parameters for circular and linear photon beam polarization, respectively.
$a\,(=\alpha_-^{\rm old}/\alpha_-)$ allows for a calibration
of the $\Lambda$ decay parameter, for which in the CLAS
publications the PDG value at that time, 
$\alpha_-^{\rm old}=0.642$, had been adopted.
We use the convention that caligraphic symbols denote random variables (RVs).
The observables ${\cal O}_{j,i}$ are  assumed independent, normally distributed RVs, ${\cal O}_{j,i}\sim {\cal N}[\mu_{j,i},\sigma_{j,i}^2]$ that take on values $O_{j,i}$. The Fierz RVs ${\cal F}^{(1,2)}_i$ take on values $f^{(1,2)}_i$ and $\mu_{j,i}$, $\sigma_{j,i}^2$ are the reported CLAS measurements.
The use of the constraints imposed by the Fierz identities to determine $a,\,l,\,c$ poses a series of statistical challenges  
 that are summarized below. The Supplemental Material~\cite{supplemental} expands on these points with several explicit derivations and numerical checks using synthetic data.

\smallskip

\begin{enumerate}[wide, leftmargin=10pt,labelwidth=!,label=\textbf{\arabic*.}]
\setlength{\itemsep}{2pt}
\setlength{\parskip}{2pt}
\setlength{\parsep}{2pt}
\item
\emph{Parameter estimates were checked to be unbiased.}
The parameters $a,l,c$ scale both the $\mu_{j,i}$ and the uncertainties $\sigma_{j,i}$, which potentially leads to biased results. This is a problem related to, but not identical to, an effect known as the d'Agostini bias~\cite{DAgostini:1993arp, Ball:2009qv}. 
\item
\emph{Unnormalised probability density functions (pdfs) were used.}
Normalization factors of likelihoods depend on the data values that, in our case, depend on $a,l,c$. This dependence is spurious~\cite{Roe:2015fca}. We therefore indicate the likelihoods with ``$\propto$'' in the following. Once the distribution of $a,\,l,\,c$ is determined we perform an a-posteriori normalization of the result, see Eq.~(\ref{eq:blablubb}) below.
\item 
\emph{
For the first Fierz identity, a naive guess based on Eq.~(\ref{eq:Fierz1}) of the expectation, $E[{\cal F}_i^{(1)}]=1$, is only correct in the limit $\sigma_{j,i}\to 0$.} The pdf of each summand in Eq.~(\ref{eq:fierzval}) follows a scaled, non-central $\chi^2$ distribution with $E[{\cal O}_{j,i}^2]=\mu_{j,i}^2+\sigma_{j,i}^2\neq \mu_{j,i}^2$. Although there exists no closed form for the distribution of ${\cal F}_i^{(1)}$, denoted below as $p^{(1)}(f_i^{(1)}|a,l,c)$ , the expectation value can be calculated because expectation values add. For ${\cal F}_i^{(2)}$, $E[{\cal O}_{j,i}{\cal O}_{j',i}]=\mu_{j,i}\mu_{j',i}$ with $j\neq j'$ and there is no such shift so that the Fierz identity reads $E[{\cal F}_i^{(2)}]=0$.

For each kinematic point $i$, we obtain 
\begin{align}
p_i^{(12)}({\mathfrak O}_{i}|a{,}l{,}c)
&{~\propto~}
p^{(1)}(f^{(1)}_i{=}\Delta f_i|a{,}l{,}c)
\nonumber \\ &{~\times~}
p^{(2)}(f^{(2)}_i{=}0|a{,}l{,}c),
\label{finonlin}
\end{align}
where ${\mathfrak O}_i=\cup_{j=1}^7 {\cal O}_{j,i}$ symbolizes the data set at point $i$. Here, $\Delta f_i\neq 1$ is the $a,\,l,\,c$-dependent expectation value for $f_i^{(1)}$ that corresponds to the best fulfillment of the first Fierz identity (see Supplemental Material~\cite{supplemental} for an explicit expression).
As there is no closed form for the distributions of the Fierz values, they can be estimated by sampling: For fixed $a,l,c$, 
Fierz values $f_{i}^{(1,2)}$ are calculated from random samples  of the observables ${\cal O}_{j,i}$. Then, those $f_{i}^{(1,2)}$ that are located in a small region around $\Delta f_i$ and $0$ are counted, for Fierz identity 1 and 2, respectively. This procedure is repeated in a scan of the whole $a,\,l,\,c$ space.
\item
\emph{A Gaussian likelihood can be used for each point $i$.}
We found that the non-linearities of the problem are small for this particular case as discussed in the Supplemental Material~\cite{supplemental}, which allows us to approximate 
\begin{equation}
p_i^{(12)}({\mathfrak O}_i|a,l,c)
\propto
\exp \left[-\left( \dfrac{\mu_{f_i^{(1)}} - 1}{\sigma_{{\cal F}_i^{(1)}}} \right)^2
-\left( \dfrac{\mu_{f_i^{(2)}}}{\sigma_{{\cal F}_i^{(2)}}} \right)^2\right]
\ ,
\label{gaussapp}
\end{equation}
where the $\mu_{f_i^{(1,2)}}$ equal the right-hand sides of Eqs.~(\ref{eq:fierzval},\ref{eq:fierzval2}) with the ${\cal O}_{j,i}$ replaced by their means $\mu_{j,i}$ (i.e.\,the measured
central values reported in the literature), and
 expressions for $\sigma_{{\cal F}_i^{(1,2)}}$ given in the Supplemental Material~\cite{supplemental}. This probability is thus an expression of how far away from the Fierz constraints the combination of the observables $j$ at kinematic point $i$ is. 
 \end{enumerate}

 \smallskip

As data for different energies and scattering angles are independent, the combined likelihood can be written as the product
\begin{equation}
{\cal P}({\mathfrak O}|a,l,c)=\frac{1}{Z}\prod_{i=1}^n p_i^{(12)}({\mathfrak O}_i|a,l,c) \ ,
\label{eq:blablubb}
\end{equation}
where ${\mathfrak O}=\cup_{i=1}^n {\mathfrak O}_i$ symbolizes the entire data set and $Z$ is the normalization constant obtained by integrating ${\cal P}({\mathfrak O}|a,l,c)$ over the $a,\,l,\,c$ space (see item {\bf 2.}).

Even with the two Fierz identities as constraints, $a,l$, and $c$ are highly correlated,  and priors on $P_C^{\gamma}$ and $P_L^{\gamma}$ are required. Systematic uncertainties in the experiments are quoted as numbers, which we denote as $\delta_C$ and $\delta_L$, but there is no universal prescription to code this information as a pdf. To check the robustness of the method we used four different priors ${\cal P}(l,c)$: 1) Gaussian: $l,c \sim {\cal N} (1,\delta_{l,c}^2)$; 2) Uniform: $l,c \sim {\cal U} (1-\delta_{l,c},1+\delta_{l,c})$; 3) Double Uniform: $l,c \sim {\cal U} (1-2\delta_{l,c},1+2\delta_{l,c})$; and 4) Fixed: $l=c=1$. We take $\delta_l=0.05$ and $\delta_c=0.02$ as representative values, according to the systematic errors estimated in Refs.\,\cite{bradford_first_2007,paterson_photoproduction_2016}. ${\cal U}$ represents a uniform pdf. The posterior density is 
\begin{equation}
{\cal P}(a,l,c|{\mathfrak O})\propto  {\cal P}({\mathfrak O}|a,l,c){\cal P}(l,c) \ .
\label{eq:posterior}
\end{equation}
The posteriors corresponding to the choice of priors were explored using a Markov Chain Monte Carlo (MCMC) implementation (\emph{emcee} \cite{2013PASP..125..306F}).
As there were only three parameters to be determined we were also able to scan directly across the parameters $a$, $c$ and $l$ to validate the results of the MCMC. The results for $\alpha_-$ were obtained by marginalizing over $l$ and $c$.  Both methods were checked by applying them to synthetic data that had been scaled appropriately by a ``wrong'' value of $\aw$. 


\section{Results}

\begin{figure}[t]
\includegraphics[width=0.99\linewidth]{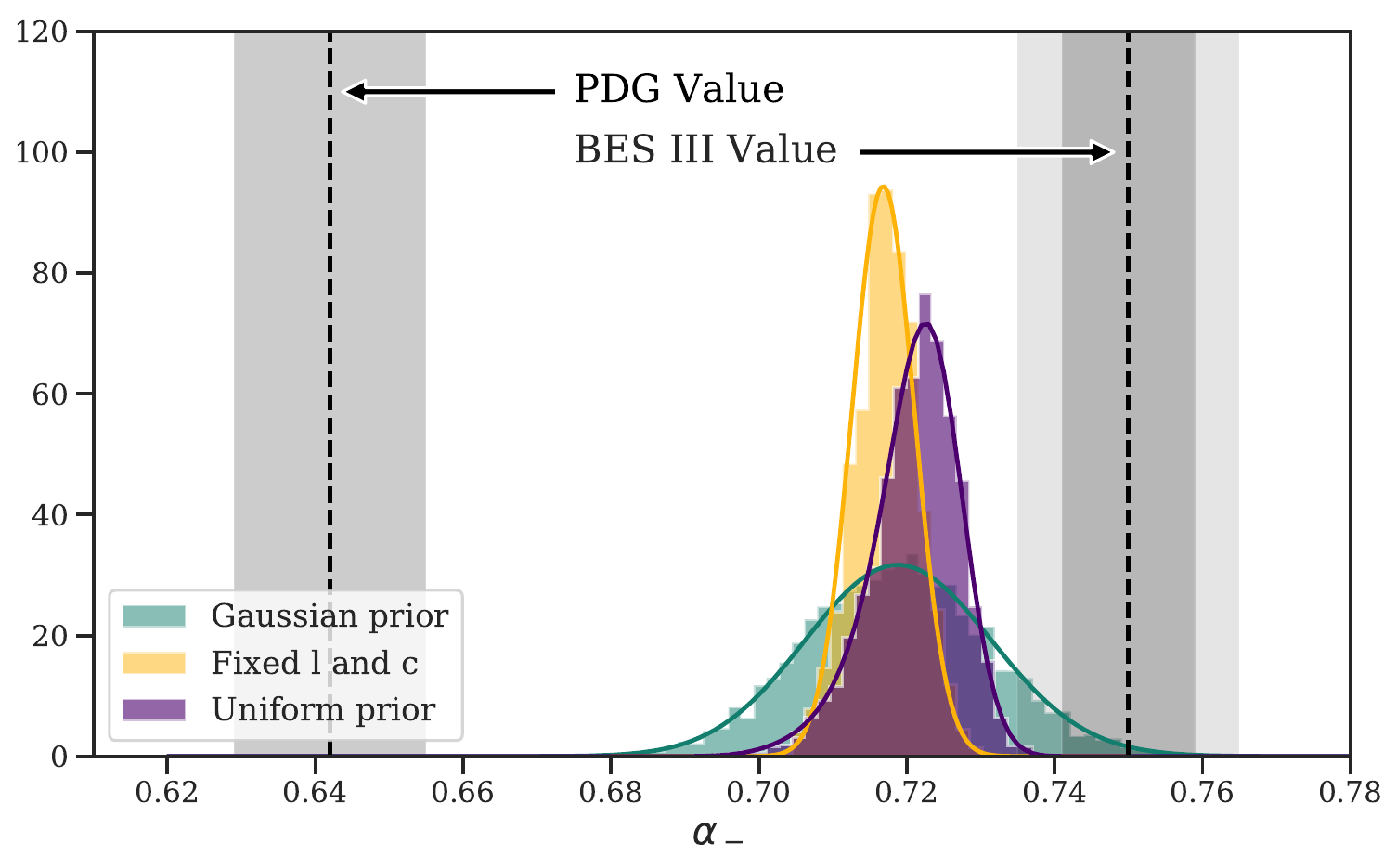}
\caption{\label{fig:posteriors}Posterior densities for $\alpha_-$, given different priors for the beam polarization calibration constants $P_C^{\gamma}$ and $P_L^{\gamma}$. The histograms show the result of the MCMC sampling of the marginalized posterior densities while the solid lines represent a direct scan of the posteriors. For clarity, the results corresponding to the double width uniform priors for $P_C^{\gamma}$ and $P_L^{\gamma}$ are omitted. Dark grey vertical bands represent statistical uncertainty; the additional light grey bands on the BESIII result represent systematic uncertainty.}
\end{figure}

The results for the marginalized posteriors for $\alpha_-$ with the measured CLAS data are depicted in Fig.~\ref{fig:posteriors}, and the mean and standard deviation of the marginalized pdfs are reported in Tab.~\ref{tab:results}. 

The means of the posteriors are all consistent with each other. Whilst this is not an exhaustive sensitivity check, the range of priors chosen reflects quite different assumptions. This therefore suggests that the estimated value for $\aw$ does not dependent sensitively on the choice of prior.

The Gaussian priors for $c$ and $l$ give unrealistic mean values of $c$ and $l$ in the posterior pdf that are 3-4 standard deviations from 1.0, their nominal values. This is possible since a normal distribution is technically non-zero over an infinite domain. Results reported by experiments imply that the range of values defined by the quoted systematic uncertainties should contain the possible values of calibration parameters with high probability, without specifying the form of a pdf. Whilst normal pdfs are often assumed for systematic uncertainty they are perhaps not appropriate in this case.

The use of uniform pdfs as priors for $P_C^{\gamma}$ and $P_L^{\gamma}$ represents another extreme, where the implication is that the true values \emph{must} lie within a given range. We take two variants: a uniform range defined by the size of the systematic uncertainties, and a uniform distribution of double this range. 
A final extreme assumption is that there is no systematic error, and that $c=l=1$. 

We make the assumption that the uniform prior for $c$ and $l$ between the quoted systematic uncertainties represents the most realistic assumption, so we quote the mean value of this variant (0.721) as our result, together with the standard deviation (0.006) of the pdf of $\aw$ as the statistical uncertainty, and a systematic uncertainty of $\pm$ half the range of values $\nicefrac{1}{2}(0.727 - 0.717)=0.005$. 
We denote this value by $\alpha^{\rm CLAS}_-$ below. 

The Supplemental Material~\cite{supplemental} provides a more detailed representation of the results in $a,l,c$ space.

\subsection{Refits with the J\"ulich-Bonn model}
To cross-check the results obtained in the previous section and to estimate the impact of a new value of $\alpha_-$ in calculations that employ data such as the ones from Refs.~\cite{bradford_first_2007, mccracken_differential_2010, paterson_photoproduction_2016} as input, we use the J\"ulich-Bonn (J\"uBo) framework. This dynamical coupled-channel approach is one framework among others~\cite{Hunt:2018mrt, Anisovich:2017bsk, Kamano:2016bgm, Cao:2013psa, Maxwell:2012zz, DeCruz:2011xi} that aim to extract the nucleon resonance spectrum from kaon photoproduction, often in a combined analysis of pion- and photon-induced hadronic scattering processes.  
In the J\"uBo approach, the Fierz identities are fulfilled by construction. A detailed description of the model can be found in Refs.~\cite{Ronchen:2012eg} and \cite{Ronchen:2014cna}; the photoproduction data of the $\eta p$ and $K^+\Lambda$ final states were included recently~\cite{Ronchen:2015vfa, Ronchen:2018ury}, among them the measurements of the differential cross section and several polarization observables in $K\Lambda$ photoproduction by the CLAS Collaboration~\cite{bradford_first_2007, mccracken_differential_2010, paterson_photoproduction_2016}.

\begin{table}[t]
    \begin{ruledtabular}
    \begin{tabular}{p{1.6cm}ll} 
    Source & Value~(stat)~(sys) & 
    Prior Assumption $c$, $l$ \\ 
    \colrule
    \addlinespace[0.5em]
    PDG'18~\cite{Tanabashi:2018oca} & 0.642 (13) \\
    BES III~\cite{Ablikim:2018zay}   & 0.750 (9) (4) \\ 
    \addlinespace[0.5em]
    \colrule
    \addlinespace[0.5em]
    Analysis & 0.719 (13) 
    & ${\cal N}(1.0,0.02^2)$, ${\cal N}(1.0,0.05^2)$ \\
    of CLAS & 0.721 (6) ($\star$)
    & ${\cal U}(0.98,1.02)$, ${\cal U}(0.95,1.05)$ \\
    data& 0.727 (7) 
    & ${\cal U}(0.96,1.04)$, ${\cal U}(0.90,1.10)$ \\
    & 0.717 (4) 
    & Both fixed at 1.0 \\
    \addlinespace[0.5em]
    & 0.721 (6) (5)  
    & Summary of our result \\
    \end{tabular}
    \end{ruledtabular}
    \caption{\label{tab:results}Summary of results. The result marked ($\star$) represents the most realistic prior on $P_C^{\gamma}$ and $P_L^{\gamma}$.}
\end{table}

In order to estimate the impact of a different value for $\alpha_-$ within the J\"uBo model, the polarization observables $T$, $O_x$ and $O_z$ from Ref.~\cite{paterson_photoproduction_2016}, $C_{x}$ and $C_{z}$ from Ref.~\citep{bradford_first_2007} and $P$ from Ref.~\cite{mccracken_differential_2010} are scaled by this value, i.e. multiplied by $(\alpha^{\rm old}_-/\alpha^{\rm BESIII}_-)$ or by $(\alpha^{\rm old}_-/\alpha^{\rm CLAS}_-)$ and a refit of a  subspace of free parameters of the model is performed. The data included in the refit are limited to those that are contained in the energy range defined by the measurement in \cite{paterson_photoproduction_2016}. Note that also the statistical data errors entering the $\chi^2$ are scaled. 

In addition, we also perform a refit of the unscaled data. This is necessary because the solution J\"uBo2017 \cite{Ronchen:2018ury}, which is the starting point for the refits, represents the minimum of the global coupled-channels fit including all 48,000 data points from different reactions. A refit considering only the unscaled data listed in Tab.~\ref{tab:chi2} provides a valid point of comparison for the fit to the scaled data. We vary only parameters of the non-pole polynomials~\cite{Ronchen:2014cna} that couple to the $K\Lambda$ final state, which amounts to 73 fit parameters. They are adjusted to the data in a $\chi^2$ minimization using MINUIT on the JURECA supercomputer at the J\"ulich Supercomputing Centre~\cite{jureca}. In all three fits identical fitting strategies are applied. 

\begin{table}[t]
\centering
\renewcommand{\arraystretch}{1.}
\begin{ruledtabular}
\begin{tabular}{rrcc}
Observable &   \multicolumn{3}{c}{$\chi^2/n$ (Refits)} \\
(\# data points)& $\alpha_-=$ 0.642 & 0.75 & 0.721 \\ \hline
    \addlinespace[0.2em]
$d\sigma/d\Omega$ (421)  \cite{mccracken_differential_2010}  & 1.11 & 1.03 & 0.95\\
$\Sigma$ (314) \cite{paterson_photoproduction_2016}  &2.55 & 2.61 & 2.56\\
$T$ (314) \cite{paterson_photoproduction_2016}&   1.75 & 1.74 &1.69\\
$P$ (410) \cite{mccracken_differential_2010}& 1.84 & 1.66 & 1.62 \\
$C_{x}$ (82) \citep{bradford_first_2007}&  2.15  & 1.72& 1.34 \\
$C_{z}$ (85) \citep{bradford_first_2007}& 1.58 &1.83 & 1.62\\
$O_x$ (314) \cite{paterson_photoproduction_2016}&  1.44 &1.53 & 1.51\\
$O_z$ (314) \cite{paterson_photoproduction_2016}& 1.34&1.58 & 1.49
\\ \hline
    \addlinespace[0.2em]
all (2254) &  1.67 &1.66 & 1.59  
\end{tabular}
\end{ruledtabular}
\caption{$\chi^2/$data point of the J\"ulich-Bonn refits for different values of $\alpha_-$. The value of $\alpha_-=\alpha_-^{\text{old}}=0.642$ corresponds to the refit to unscaled data,  $\alpha_-=0.75$ correponds to the BES-III result~\cite{Ablikim:2018zay} and $\alpha_-=0.721$ uses the data-driven result of this study as input for the refit.
} 
\label{tab:chi2}
\end{table}

The results are shown in Table~\ref{tab:chi2}. The best $\chi^2$ is obtained for the data scaled by $\alpha^{\rm CLAS}_-$ as determined in this study, while the refit to the data scaled by $\alpha^{\rm BESIII}_-$ returns a similar $\chi^2$ to the fit to the unscaled data ($\alpha^{\rm old}_-=0.642$). Both are significantly worse than  $\alpha^{\rm CLAS}_-$ which corroborates our independent result. As a caveat, the best $\chi^2/n$ itself (1.59) is still too large which suggests that for a more quantitative comparison $l$ and $c$ should also be varied as before to allow for more systematic uncertainties, or that the model parameterization itself is not flexible enough. 


\section{Conclusions}
The decay parameter $\aw$ of the $\Lambda$ is a fundamental physical constant that is used to obtain polarization information from reactions in which the parity-violating weak decay $\Lambda \rightarrow p \pi^-$ occurs. Its value has recently been thrown into dispute by a new measurement, thereby affecting all results that rely on it. We have made an independent estimate of this quantity by combining an ensemble of observables from kaon photoproduction measured at CLAS with constraints set by Fierz identities. Our value of 0.721 $\pm$ 0.006 (statistical) $\pm$ 0.005 (systematic), clearly favours the new BESIII 
result of 0.750 $\pm$ 0.009 $\pm$ 0.004 over the previous PDG value of 
0.642 $\pm$ 0.013, though it differs manifestly from the former as well.

In view of that, it is clear that past results which involve
the $\Lambda$ decay parameter should be revisited to ensure that the derived quantities are in line with the new and larger reference value of $\aw$,
bearing in mind the remaining uncertainty.
This applies to data from all experiments where the polarization of the $\Lambda$ or $\Xi$ baryon was measured. As a consequence, 
phenomenological analyses of those data performed in searches for (new) excited baryons and their properties should also be updated. 



\bigskip

\begin{acknowledgments}
This work was supported by: the United Kingdom's Science and Technology Facilities Council (STFC) from grant number ST/P004458/1;
M.D. acknowledges support by the National Science Foundation (grant no. PHY-1452055) and by the U.S. Department of Energy, Office of Science, Office of Nuclear Physics under contract no. DE-AC05-06OR23177 and grant no. DE-SC0016582. This work is supported in part by DFG and NSFC through funds provided to the Sino-German CRC 110 ``Symmetry and the Emergence of Structure in QCD" (NSFC Grant No. 11621131001, DFG Grant No. TRR110) and R.M-S acknowledges EPSRC grants  EP/M01326X/1 \& EP/R018634/1. The authors gratefully acknowledge the computing time granted through JARA-HPC on the supercomputer JURECA at Forschungszentrum Jülich. We thank Dr H. Nickisch for helpful advice on the implementation of coupled heteroskedastic GP models in his GPML toolbox.\footnote{\url{https://gitlab.com/hnickisch/gpml-matlab}} We also thank Ulf-G.~Meißner for his comments on the draft of this article.
\end{acknowledgments}


%


\end{document}